\documentclass[fleqn,usenatbib]{mnras}
\usepackage{newtxtext,newtxmath}

\usepackage[T1]{fontenc}

\DeclareRobustCommand{\VAN}[3]{#2}
\let\VANthebibliography\thebibliography
\def\thebibliography{\DeclareRobustCommand{\VAN}[3]{##3}\VANthebibliography}

\usepackage{graphicx}	
\usepackage{amsmath}	
\usepackage[figuresright]{rotating}
\usepackage{afterpage}
\usepackage{caption}
\usepackage{xcolor}
\colorlet{RED}{red}
\usepackage{animate}

\newcommand\yep{Yep et al., in preparation}

\newcommand\nodata{$\cdots$}

\title[Collision of Two Stellar Associations]{Collision of Two Stellar Associations in the Nearby Gum Nebula}

\author[Yep \& White]{Alexandra C. Yep,$^{1}$\thanks{E-mail: ayep@astro.gsu.edu (ACY)}$^{2}$\thanks{E-mail: ayep@agnesscott.edu (ACY)} Russel J. White$^{1}$
\\
$^{1}$Astronomy Department, Georgia State University, 25 Park Place, Atlanta, GA 30333, USA\\
$^{2}$Physics and Astronomy Department, Agnes Scott College, Bradley Observatory, Decatur, GA 30030, USA\\
}

\date{Accepted 2021 December 18. Received 2021 December 18; in original form 2021 April 30}

\pubyear{2021}

\begin{document}
\label{firstpage}
\pagerange{\pageref{firstpage}--\pageref{lastpage}}
\maketitle

\begin{abstract}
Based on \textit{Gaia} DR2 data and new CHIRON radial velocities, we have discovered that two nearby stellar associations UPK 535 (318.08 $\pm$ 0.29 pc, $25^{+15}_{-10}$ Myr, 174 stars) and Yep 3 (339.54 $\pm$ 0.25 pc , $45^{+55}_{-20}$ Myr, 297 stars) in the Gum Nebula have recently collided. We project stars' current positions, motions, and measurement uncertainties backward and forward through time in a 10,000-trial Monte Carlo simulation. On average, the associations' centres of mass come within 18.89 $\pm$ 0.73 pc of each other 0.84 $\pm$ 0.03 Myr ago. A mode of 54 $\pm$ 7 close ($<$1 pc) stellar encounters occur during the collision. We cannot predict specific star-star close encounters with our current $\sim$7.6 pc distance precision and 21.5-per-cent-complete radial velocity sample. Never the less, we find that two stars in UPK 535 and two stars in Yep 3 undergo a nonspecific close encounter in $>$70 per cent of trials and multiple close encounters in $\sim$30 per cent. On average, the closest approach of any two stars is 0.13 $\pm$ 0.06 pc, or 27,000 $\pm$ 12,000 au. With impulse-tracing values up to $2.7^{+3.1}_{-1.1}$ M$_{\odot}$ pc$^{-2}$ km$^{-1}$ s, such close encounters could perturb stars' Oort cloud comets (if present), cause heavy bombardment events for exoplanets (if present), and reshape solar system architectures. Finally, an expansion of our simulation suggests other associations in the region are also interacting. Association collisions may be commonplace, at least in the Gum Nebula straddling the Galactic plane, and may spur solar system evolution more than previously recognized.
\end{abstract}




\begin{figure}
\centering
\includegraphics[scale=0.995]{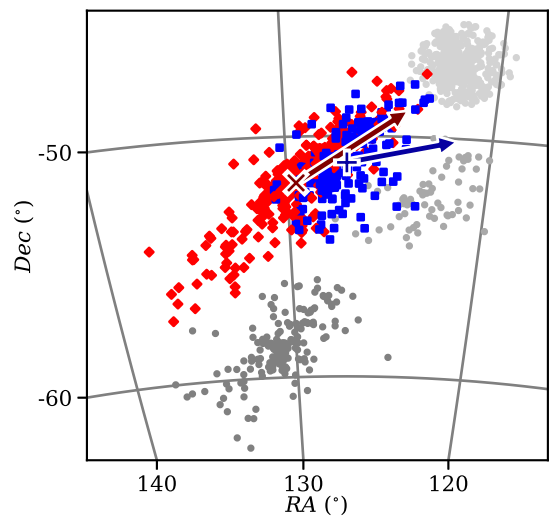}
\caption{The sky positions of five associations in the Gum Nebula. Associations UPK 535 (blue squares) at 318.08 $\pm$ 0.29 pc and Yep 3 (red diamonds) at 339.54 $\pm$ 0.25 pc have elongated shapes and spatially overlap. Their centres of mass are marked by a dark blue \textit{+} and a dark red \textit{x}, respectively. Centre-of-mass motions over 1 Myr are marked by dark blue and dark red arrows. UPK 535 and Yep 3 are near other associations UPK 545 at $326.80\pm0.24$ pc (dark grey circles, south), UPK 533 at $344.08\pm0.46$ pc (grey circles, west), and Pozzo 1 at $346.75\pm0.23$ pc (light grey circles, northwest).}
\label{CCmaptoday}
\end{figure}

\section{Introduction}

Because a cloud of molecular gas fragments as it collapses, most stars form in clusters or associations \citep{lada,krum}. Modern clusters and associations consist of hundreds or thousands of stars of similar age and composition, all in close proximity to each other and moving through space together. Clusters are gravitationally bound and can remain coherent for $>$100 Myr, while associations, sparser than clusters, are gravitationally unbound and naturally drift apart in 10 -- 100 Myr \citep{mathieu,lada,krum,kouncov}. With the advent of \textit{Gaia} DR2 \citep{gaia2A,Gaia2018} and supervised and unsupervised machine learning efforts from \citet{kouncov}, \citet{cant18}, \citet{gin}, and others, two to three thousand open clusters and associations are known, and the census is far from complete \citep{mor,gin}.


In regions where many clusters and associations form, like the Gum Nebula in the Galactic plane \citep{gum}, clusters and associations may dynamically interact and possibly collide. Cluster and association collisions have not been widely considered in astronomy. However, any discovered colliding clusters or associations could serve as ideal laboratories for studying close stellar encounters and their effects on solar systems. Short of collisions, spatially overlapping clusters and associations have been detected, including NGC 1750 and NGC 1758 \citep{galadi}, two components of $\sigma$ Ori \citep{jeff06}, two components of R136 \citep{sabbi2}, and two components of $\gamma$ Vel OB2 \citep{jeff,sac}. Using \textit{Gaia} DR2 data, \citet{wp} found two mass-separated components of NGC 6530, where the motions of high-mass stars differ from those of low-mass stars. Stellar overdensity within a cluster can also increase stellar interactions, possibly causing hot jupiters (\citealp{wintplanet}; \citealp{longmore}; see \citealp{adib} for caveat on cluster age) and changing the orbits of superearths \citep{rodet}.

Approaching this topic from another angle, astronomers have searched for and identified stars that have or will pass close to the Sun. These have the potential to perturb Oort cloud comets into inner orbits, potentially resulting in heavy-bombardment or mass-extinction events \citep{weiss,yeo,fengbj,bj2}. In short, despite the vast emptiness of space, clusters and associations can and do pass near each other, and stars in close proximity to other stars can affect each other's planets.


Our paper introduces a first-ever kinematic case study of two distinct stellar associations in the process of colliding (see Figure \ref{CCmaptoday}). The associations are UPK 535 and Yep 3 in the Gum Nebula \citep{gum,sim,cant,kouncov}. Whereas the overlapping populations of $\sigma$ Ori, R136, $\gamma$ Vel OB2, and NGC 6530 are each believed to be evolving components of a single cluster \citep{jeff06,sabbi2,jeff,wp}, with distinguishable but similar space motions, UPK 535 and Yep 3 exhibit disparate space motions and spatially overlap as a result of a chance encounter. Their collision could shed light on how association interactions affect association structure, association dispersal, and solar systems.

In \S\ref{ID}, we identify association members. In \S\ref{spec}, we derive spectral types, masses, and radial velocities from spectra of 95 members and \textit{Gaia} DR2 data of all members. In \S\ref{discussion}, we analyse kinematics, a linear-motion Monte Carlo simulation, other associations in the vicinity of UPK 535 and Yep 3, and association kinetic and potential energies. Finally, in \S\ref{summ}, we summarize our results.

\section{Identification of Stellar Associations} \label{ID}


\begin{table}
	\centering
    \caption{Position and motion cuts for isolating the associations UPK 535 and Yep 3 from \textit{Gaia} DR2 data. We also impose error cuts $<0.1$ mas in parallax and $<0.16$ mas yr$^{-1}$ in proper motion.}
	\label{CCpars}
\resizebox{1.02\columnwidth}{!}{
\hspace{-15pt}
\begin{tabular}{lccccc} 
\hline
Assn. & $\mathit{RA}$ & $\mathit{Dec}$ & $d$ & $\mu_{\alpha}$ & $\mu_{\delta}$ \\  
Name & ($^{\circ}$) & ($^{\circ}$) & (pc) & (mas yr$^{-1}$) & (mas yr$^{-1}$) \\ 
\hline
UPK 535 & 123.0 -- 131.0 & -55.0 -- -47.5 & 290 -- 350 & -14.5 -- -11.5 & 1.1 -- 4.1 \\ 
Yep 3 & 126.3 -- 134.0 & -55.0 -- -49.5 & 320 -- 370 & -14.5 -- -11.5 & 9.0 -- 12.0 \\ 
\hline
\end{tabular}}
\end{table}



We identified the stellar associations UPK 535 and Yep 3 as part of an ongoing study of stars associated with cometary globules in the Gum Nebula \citep{yep}. We developed a \textsc{python} code called Cluster Finder that facilitates empirical detection of spatially compact groups of stars with consistent distances and motions.\footnote{\url{https://github.com/alexandrayep/Cluster\_Finder}} All our searches begin with a 2 -- 4$^{\circ}$-radius \textit{Gaia} DR2 field within the Gum Nebula. We then administer parallax- and proper-motion cuts. Based on error analysis by \citet{luri}, we impose parallax measurement error $<0.1$ mas and proper motion measurement error $<0.16$ mas yr$^{-1}$ to eliminate sources with poor astrometry. We fine-tune cuts to favor probable association membership over comprehensiveness, narrowing the cuts until the association is clearly visible and the roughly uniform distribution of stars surrounding the association dwindles to near zero. Lastly, we crop right ascension and declination to the association edges. Final distances are from \citet{bj} (see Table \ref{CCpars}). Using this technique, we found eight associations throughout the Gum Nebula, two of which spatially overlap each other (see Figures \ref{CCmaptoday} and \ref{CC_Histograms}).

These two associations have been previously identified as UPK 535 in \citet{sim} and \citet{cant} and, loosely, as Theia 120 in \citet{kouncov}. \citet{cant} and \citet{kouncov} utilize primarily \textit{Gaia} DR2 parallaxes and proper motions to identify association members. We combine our membership lists with these other catalogues', imposing our listed parameter cuts (see Table \ref{CCpars}) with slight expansions of $\pm5$ pc in distance range and $\pm0.1$ mas yr$^{-1}$ in proper motion ranges. We reject stars with distances farther than 500 pc and distance uncertainties $>$50 pc. Thus for UPK 535 in \citet{cant} we include 86 stars that appear in both our membership lists, 30 stars that appear in only their list, and 58 stars that appear in only our list, for a total of 174 members of UPK 535.

Theia 120 in \citet{kouncov} appears spuriously large, with 1633 members extending from $\mathit{RA}$ 120$^{\circ}$ to 240$^{\circ}$. We consider only the clustered portion west of $\mathit{RA}$ 148$^{\circ}$, with our cuts applied as aforementioned. We include 111 stars that appear in both our membership lists, 153 that appear in only their list, and 33 that appear in only our list, for a total of 297 members of Yep 3. The Yep 3 association may include as many as 227 additional members of Theia 120, but these fall outside our membership-probability-favoring cuts, are thus less likely to all be members, and are not considered in this study.






\section{Optical Spectra and Stellar Properties} \label{spec}

\subsection{Observations and Data Reduction} \label{obs} 

From 2018 October 22 to 2020 March 10, we observed 36 stars in UPK 535 and 59 stars in Yep 3 for at least 1 epoch each with the CHIRON spectrograph in queue-scheduled fibre mode (\citealp{toko}; \citealp{leo}). The instrument covers 4500 -- 8500 \AA\ at a resolving power $R \approx 25,000$, corresponding to a velocity resolution of 12 km s$^{-1}$. Guided by association candidate members' absolute \textit{Gaia} blue magnitudes $M_{BP}$ vs.\ \textit{Gaia} colours $BP-RP$ (see Figure \ref{CC_Isochrones}), we avoided stars above the apparent single-star main sequences and prioritized photometrically single F-, G-, and K-type stars to measure radial velocities. We aimed for a signal-to-noise ratio of 10 -- 30, which is sufficient to determine spectral types and radial velocities. We limited exposure times for magnitudes $V > 9.3$ mag to 1200 s in the interest of surveying all cool single $V<13.5$ mag stars in a timely manner.




CHIRON echelle spectra are reduced by the CHIRON instrumentation team using an IDL script (\citealp{leo}). Because the instrument's temperature, fibre illumination, and order position on the CCD are very stable \citep{toko}, we are able to normalize all spectra by dividing out a blaze function derived order-by-order from the fairly featureless, slow-rotating A3V-type star HD 11753. We focus on 30 orders that are mostly free of telluric features and strongly pressure-broadened lines.

A comprehensive analysis of these spectra is being assembled as part of a larger population study of stars and associations in the Gum Nebula. Here we present a summary of the methods used to determine the spectral types, extinctions, masses, and radial velocities for members of the UPK 535 and Yep 3 associations. A more thorough methodology will be presented in \yep.

\subsection{Spectral Types, Colour Excesses, and Extinctions} \label{spectra}

Spectral types of the 95 spectroscopically observed stars are determined via visual comparison to our \mbox{CHIRON} catalogue of well-measured, slow-rotating spectral standards.\footnote{\url{https://github.com/alexandrayep/CHIRON\_Standards}} We estimate a spectral type uncertainty of 1 subclass for most stars, up to 2 -- 3 subclasses for faint and fast-rotating stars. Spectral types range from K3.5 to B9 in UPK 535 and from K2.5 to B3 in Yep 3.



We measure the \textit{Gaia} colour excess $E(BP-RP)$ of each association by comparing spectroscopically observed stars' apparent \textit{Gaia} colours $BP-RP_{\text{obs}}$ with their spectral types' intrinsic colours $BP-RP_{\text{int,M}}$ according to the dwarf colours of \citet{mam}.\footnote{\url{http://www.pas.rochester.edu/~emamajek/EEM_dwarf_UBVIJHK_colors_Teff.txt}, version 2021.03.02} To avoid biasing results with anomalous red or blue outliers, possibly caused by a variation in local extinction, or skewed by dim or fast-rotating stars that are more difficult to classify, we measure overall association colour excesses $E(BP-RP)_{\text{assn}}$ by taking the flux-error-weighted mean of the middle quartiles of $E(BP-RP)$ values. This yields a colour excess of $0.059\pm0.022$ mag for UPK 535 and $0.036\pm0.027$ mag for Yep 3. Uncertainties are the flux standard deviations of the middle quartiles of $E(BP-RP)$ values. Since these stars are old enough ($>$10 Myr) to have lost the majority of their circumstellar material \citep{haisch}, we assume that $E(BP-RP)_{\text{assn}}$ represents reddening along line of sight. We determine all stars' intrinsic colours $(BP-RP)_{\text{int}}$ by correcting for association reddening: $(BP-RP)_{\text{int}}=(BP-RP)_{\text{obs}}-E(BP-RP)_{\text{assn}}$. Values for $(BP-RP)_{\text{int}}$ are -0.122 -- 3.286 mag for UPK 535 and -0.189 -- 3.202 mag for Yep 3 (see Table \ref{sample}).


There is no one-to-one relation between colour and extinction in $BP$ or $RP$ \citep{andrae}. There is, however, an approximate relation between extinction in \textit{Gaia} magnitude $G$ and $E(BP-RP)$ that follows from the PARSEC models: $A_G \approx 2 \cdot E(BP-RP)$ \citep{andrae}. We adopt this approximation to calculate association extinction $A_G \approx 2 \cdot E(BP-RP)_{\text{assn}} \approx$ 0.120 $\pm$ 0.043 mag for UPK 535 and 0.072 $\pm$ 0.053 for Yep 3 (see Table \ref{CG_properties_table.dat}). Overlapping at similar distances from Earth, these associations can be expected to have the same level of extinction. Their values are within 1$\sigma$ of each other. From these extinctions and \citet{bj} distances, we calculate corrected absolute magnitude $M_G$ for each star. Values for $M_G$ are 0.612 -- 11.657 mag for UPK 535 and -1.094 -- 11.834 mag for Yep 3.





\subsection{Radial Velocities} \label{rv} 

We measure stars' radial velocities $v_{\text{r}}$ via cross-correlation with our catalogue of CHIRON standards. The velocity of each star relative to a standard star is determined from the Doppler-uncertainty-weighted-mean average of the 30 orders used in the analysis. Uncertainties $\sigma _{v _{\text{r}}}$ are determined from the sample standard deviation of the orders' relative velocities, added in quadrature with the uncertainty in the standard's radial velocity. Barycentric corrections are calculated using the PyAstronomy function \textit{helcorr} \citep{pya}.\footnote{\url{https://github.com/sczesla/PyAstronomy}} Six rapidly rotating stars ($v_{\text{rot}}\sin(i) \gtrsim$ 100 km s$^{-1}$) have velocity dispersions $>$10 km s$^{-1}$ across their orders and yield spurious results; we consider their $v_{\text{r}}$ undetermined. For three Yep 3 stars we did not spectroscopically observe, we adopt \textit{Gaia} DR2 $v_{\text{r}}$. The median $\sigma _{v _{\text{r}}}$ of the two-association sample is $\sim$0.33 km s$^{-1}$, comparable to the median proper motion uncertainty 0.24 mas yr$^{-1}$ $\approx$ 0.41 km s$^{-1}$.




All stars in this study were chosen for their association-consistent properties. Inconsistent $v_{\text{r}}$ may therefore be a symptom of binarity rather than nonmembership. We mark all stars with $v_{\text{r}} >$5 km s$^{-1}$ divergent from its association's median as possible single-line binaries. Based on this criterion, our spectroscopically observed UPK 535 sample has five possible single-lined binaries, one of which is confirmed based on multiple $v_{\text{r}}$ measurements. In our spectroscopically observed Yep 3 sample, twelve stars are identified as possible single-lined binaries, and one star is visually suspected to be a double-lined spectroscopic binary.


 

The mean radial velocity of each association is calculated using the measured $v_{\text{r}}$ of candidate members in this study, excluding stars identified as possible binaries. From twenty-four stars' radial velocities, UPK 535 has an error-weighted mean radial velocity of 10.14 $\pm$ 0.06 km s$^{-1}$ with a standard deviation of 1.6 km s$^{-1}$. From thirty-eight stars' radial velocities, Yep 3 has an error-weighted mean radial velocity of 19.40 $\pm$ 0.04 km s$^{-1}$ with a standard deviation of 1.6 km s$^{-1}$.

\subsection{Masses} \label{mass}


We can estimate stars' masses using the dwarf colour relations of \citet{mam}. For the 95 spectroscopically observed stars, we interpolate masses from spectral types. Uncertainties are propagated from spectral type uncertainties and limited to $\geq$5 per cent of the stars' masses to account for uncertainties in choice of stellar model. For stars without spectra, we can interpolate masses from $(BP-RP)_{\text{int}}$ or, if $BP$ or $RP$ are unavailable, from $M_G$. We test the three methods on stars with measured spectral types, $(BP-RP)_{\text{int}}$, and $M_G$: Spectral-type-derived masses and intrinsic-colour-derived masses differ by $\pm$8 per cent on average, whereas spectral-type-derived masses and $M_G$-derived masses differ by $\pm$11 per cent on average. Accordingly, 366 stars without spectra are assigned intrinsic-colour-derived masses, with uncertainties propagated from colour uncertainties and limited to $\geq$8 per cent of their masses. Four stars in UPK 535 and 6 stars in Yep 3 that lack spectral types and colours are assigned $M_G$-derived masses, with mass uncertainties propagated from $G$- and distance uncertainties and limited to $\geq$11 per cent of their masses. Stellar masses range from 0.17 to 2.75 M$_{\odot}$ in UPK 535 and from 0.18 to 5.40 M$_{\odot}$ in Yep 3 (see Table \ref{sample}).

Our stellar mass uncertainties are statistical. Systematic uncertainties, especially because the stars are pre-main sequence, are likely larger.

\section{Discussion} \label{discussion}

\subsection{Association Properties} \label{associationprop}


\setlength\tabcolsep{7pt}
\begin{table*}
	\centering
    \caption{Association properties derived from corrected colour-magnitude diagrams, spectral types, spectroscopically measured radial velocities, and ellipsoid fits.}
	\label{CG_properties_table.dat}
\begin{tabular}{lcccccccccc} 
\hline
Assn. & No. & Age & Adopted & $E(BP-RP)$ & $A_G$ & Infrared & $M_{\text{tot}}$ & $v_{\text{r}}$ & $\rho_h$ & $t_{\text{cr}}$ \\
Name & Stars & (Myr) & [Fe/H] & (mag) & (mag) & Excess (mag) & (M$_{\odot}$) & (km s$^{-1}$) & (M$_{\odot}$ pc$^{-3}$) & (Myr) \\
\hline
UPK 535 & 174 & 25$^{+15}_{-10.}$ & 0.1 & 0.059 $\pm$ 0.022 & 0.120 $\pm$ 0.043 & 0.053 $\pm$ 0.016 & 145 $\pm$ 13 & 10.1 $\pm$ 1.6 & 0.008 & 82 \\
Yep 3 & 297 & 45$^{+55}_{-20.}$ & 0.2 & 0.036 $\pm$ 0.027 & 0.072 $\pm$ 0.053 & 0.024 $\pm$ 0.022 & 252 $\pm$ 27 & 19.4 $\pm$ 1.6 & 0.006 & 92 \\
\hline
\end{tabular}
\end{table*}
\setlength\tabcolsep{6pt}

\begin{figure}
\centering
\includegraphics{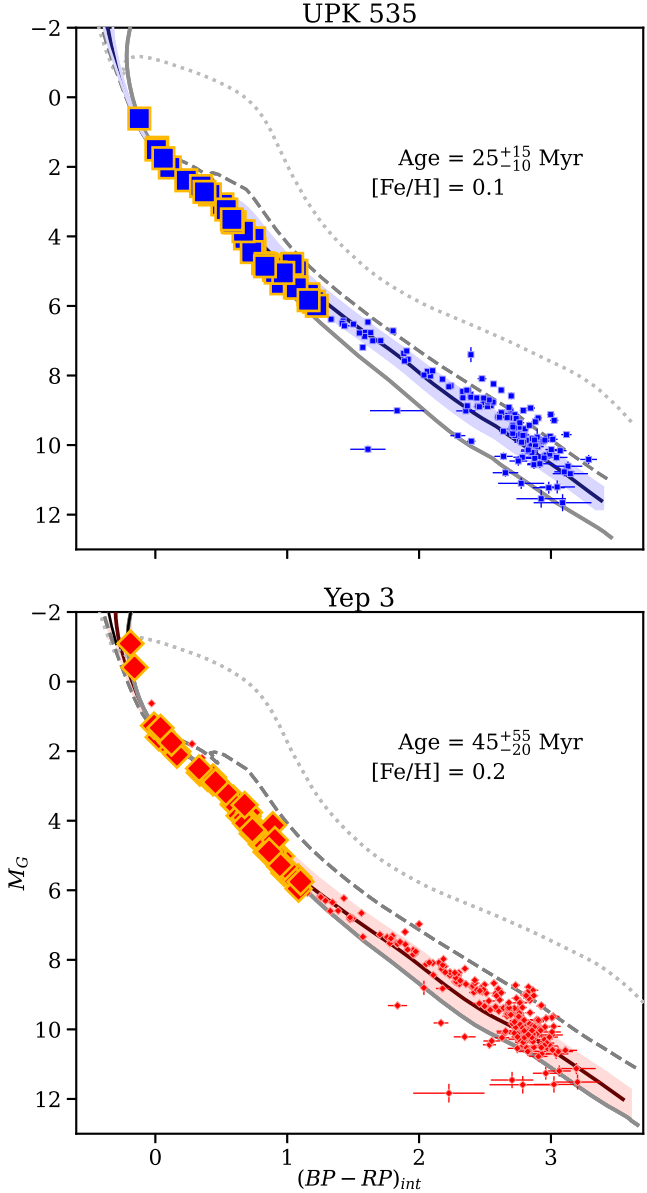}
\caption{Colour-magnitude diagrams for the associations UPK 535 (top panel) and Yep 3 (bottom panel). \textit{Gaia} colours are corrected for redenning, and absolute $G$ magnitudes are calculated using distances from \citep{bj} and corrected for extinction. Isochrones are from MESA: dotted gray is 1 Myr, dashed gray is 10 Myr, and solid gray is 100 Myr. Spectroscopically observed stars are outlined in gold. Associations UPK 535 (blue squares) and Yep 3 (red diamonds) have pre-main sequence slopes consistent with supersolar metallicities [Fe/H] = 0.1 dex and 0.2 dex, respectively, and lie roughly along isochrones of ages $25^{+15}_{-10}$ Myr (dark blue line with light blue range) and $45^{+55}_{-20}$ Myr (dark red line with light red range), respectively.}
\label{CC_Isochrones}
\end{figure}

\begin{figure}
\centering
\includegraphics{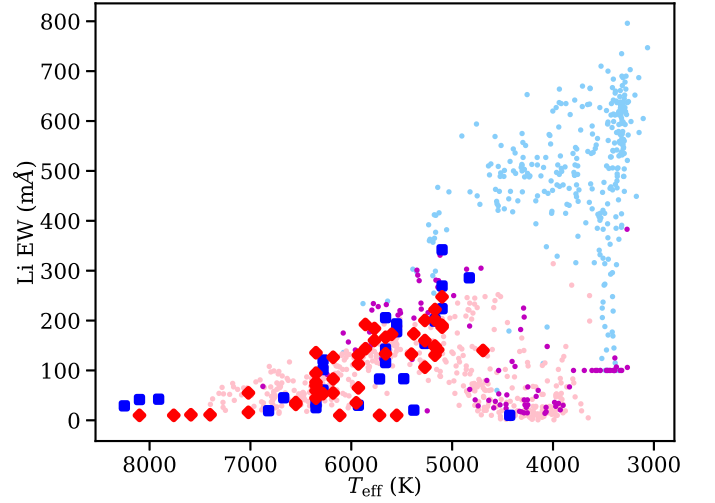}
\caption{Lithium absorption equivalent width vs.\ effective temperature. Light blue dots are for clusters and associations from \citet{guti} aged 1 -- 20 Myr, purple dots for $\sim$35 Myr, and pink dots for 100 -- 500 Myr. UPK 535 (blue squares) falls nearest the 35-Myr age sample, while Yep 3 (red diamonds) is consistent with the 100 -- 500-Myr age sample. The magnitude limit $V<13.5$ mag for our observations of UPK 535 and Yep 3 confines our sample's $T_{\text{eff}}$ to $\gtrsim$4500 K.}
\label{LiTeff}
\end{figure}

We estimate association ages by fitting MESA \citep{pax11,pax13,pax15,choi,dot16} isochrones to the extinction-corrected (see \S\ref{spectra}) single-star main sequences, as illustrated in Figure \ref{CC_Isochrones}. We find that both associations are young, with UPK 535 ($25^{+15}_{-10}$ Myr) younger than Yep 3 ($45^{+55}_{-20}$ Myr) (see Table \ref{CG_properties_table.dat}). We adopt supersolar metallicities of 0.1 dex for UPK 535 and 0.2 dex for Yep 3 because these values yield more consistent ages for stars spanning from the main sequence turnoffs to the low-mass ends, but discrepancies persist, resulting in the age uncertainties. Upper estimates are from fitting the main sequence turnoffs, while lower estimates are from fitting bright-for-their-colour low-mass stars. Such early- vs.\ late-type discrepancies are a common problem when fitting model isochrones, perhaps due to magnetism, star spots, or other difficult-to-quantify phenomena of cool stars \citep{herc,ase}, and perhaps due to blue stragglers \citep{beas}.


We buttress our isochrone results with measurements of cool stars' lithium absorption lines, prominent in cool stars when they are young. Li \textsc{I} $\lambda$6708 is visible in both associations' spectra. The feature is slightly stronger in UPK 535 (lithium equivalent width (Li EW) up to 342 m\AA) than in Yep 3 (Li EW up to 248 m\AA). We plot the associations in Li EW vs.\ effective temperature $T_{\text{eff}}$ space (see Figure \ref{LiTeff}) and compare them with empirical data from \citet{guti}. UPK 535's position falls between the 35-Myr and 100 -- 500-Myr age samples but lies closest to the 35-Myr age sample. Yep 3's position is roughly consistent with the 100 -- 500-Myr age sample. Lithium depletion is most clearly measured in low-mass stars aged $\gtrsim$100 Myr \citep{sod}, so our young associations' lithium-derived age estimates are only approximate, but they are none the less roughly consistent with the upper ends of our isochrone-derived age estimates.




The total stellar mass of each association can be calculated by summing individual stellar masses, adding the estimated mass of unobserved cool stars according to an initial mass function (IMF), and adding estimated unresolved binary companion masses until reaching a total binarity of 50 per cent.

Single stellar masses sum to 95.9 $\pm$ 7.2 M$_{\odot}$ for UPK 535 and 175 $\pm$ 19 M$_{\odot}$ for Yep 3. Uncertainties are worst-case uncertainties, summed directly.

Our samples extend down to apparent $G\sim20$ mag and are reasonably complete down to $\sim0.2$ M$_{\odot}$, or spectral type $\sim$M4V. According to the IMF of \citet{kroupa}, 48 per cent of stars are M-type stars and contribute 28 per cent of the total association mass, and 38 per cent are brown dwarfs that contribute 4.3 per cent of the total association mass. Counting all stars up to each association's largest stellar mass, we calculate that we are missing about 20 per cent of the single-star mass of UPK 535 and 12 per cent of the single-star mass of Yep 3. Completing the IMF thus adds 19.6 $\pm$ 3.9 M$_{\odot}$ to UPK 535 and 20.3 $\pm$ 4.0 M$_{\odot}$ to Yep 3, with assigned uncertainties of 20 per cent to account for uncertainties in choice of IMF and the edge of our samples' spectral type completeness.

To account for binaries, we double the mass of the one double-line binary star in Yep 3, and we approximate each confirmed or suspected single-line binary star's companion mass as half the mass of the primary (see \S \ref{rv} for binary criteria). This adds 6.7 $\pm$ 0.4 M$_{\odot}$ to UPK 535 from five companion stars and 17.9 $\pm$ 1.4 M$_{\odot}$ to Yep 3 from thirteen companion stars. If the associations have 50 per cent binarity, and if companion stars each possess half the mass of their primaries, randomly assigned unresolved binarity over 10,000 trials adds 22.3 $\pm$ 1.5 M$_{\odot}$ to UPK 535 and 39.2 $\pm$ 2.2 M$_{\odot}$ to Yep 3.

Summing all masses and uncertainties, the association total stellar mass $M_{\text{tot}}$ with worst-case uncertainty is 145 $\pm$ 13 M$_{\odot}$ for UPK 535 and 252 $\pm$ 27 M$_{\odot}$ for Yep 3 (see Table \ref{CG_properties_table.dat}). Both associations are older than 5 Myr, so we assume their natal molecular gas has fully dispersed \citep{lada}. Thus the associations' total masses are assumed equal to their association stellar masses. Total stellar mass uncertainties are statistical; systematic uncertainties are likely higher.

\begin{figure}
\centering
\includegraphics[scale=1]{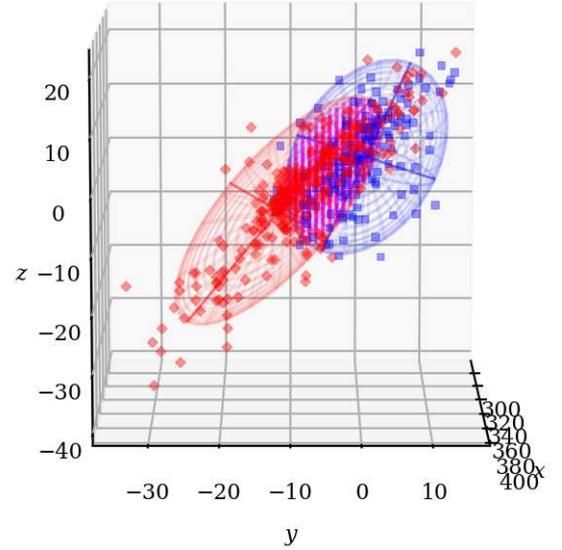}
\caption{Animated rotating view of UPK 535 (blue squares) and Yep 3 (red diamonds) today. The ellipsoid volume overlap region (magenta) contains 50 UPK 535 stars and 43 Yep 3 stars. Animation is also available in online resources.}
\label{anfigrottoday}
\end{figure}

Both associations are distinctly nonspherical (see Figure \ref{CCmaptoday}). They are elongated in the southeast-northwest direction, especially Yep 3. Because the elongation stretches roughly along the Galactic plane, tidal disruption could be at least partly responsible \citep{chenT}. To examine the spatial distribution of each association, we project their stars' 3-D positions into $xyz$ space, with $\mathit{RA}$ $\sim$ $y$, $\mathit{Dec}$ $\sim$ $z$, and distance $\sim$ $x$. We set the origin at the median position of UPK 535. Median distance uncertainties $\sigma_d$ $\sim$ 7.6 pc are significantly larger than median spatial uncertainties $\sigma_{RA} \approx \sigma_{Dec} \sim$ 0.01 au. This anisotropy in uncertainties artificially stretches the associations in the radial direction. The stretch is hidden when distance is projected into the depth direction $x$ but is revealed in other projections (see Figure \ref{anfigrottoday}).


Referencing stars' $xyz$ positions and estimated stellar masses, including binary companion masses for identified potential binary stars (see \S\ref{mass}), we determine each association's centre of mass. As an approximation of association size, we take the median of stars' distances from each association's centre of mass. These median radial extents are 14.6 pc for UPK 535 and 16.3 pc for Yep 3.


For volume, density, and stellar crossing time, we fit an ellipsoid to each association using the \textsc{python} code \textsc{ellipsoid},\footnote{Dr. Imelfort, \url{https://github.com/minillinim/ellipsoid}} with fit tolerance set to 0.15. This captures the overall morphology of each association. The initial ellipsoid for UPK 535 contains 89.1 per cent of stars, for Yep 3, 91.9 per cent. We define a half-mass ellipsoid by dilating the initial ellipsoid until it contains half the measured mass of the association (i.e.\ accounting for identified possible binaries, but excluding unidentified binaries and IMF additions).

To account for anisotropically large uncertainties in distance, we subtract out each association's projected median distance uncertainty in quadrature from its three ellipsoidal axes. We then compute the volume of the half-mass ellipsoid and calculate stellar density $\rho_h$, equal to 0.008 M$_{\odot}$ pc$^{-3}$ for UPK 535 and 0.006 M$_{\odot}$ pc$^{-3}$ for Yep 3. These densities are an order of magnitude lower than the local field star density of 0.09 pc$^{-3}$ \citep{todd} but are reasonable for stellar associations \citep{mor}. The densities imply stellar crossing times $t_{\text{cr}}\approx1/2\:(G\,\rho_h)^{-1/2} \approx$ 82 Myr for UPK 535 and 92 Myr for Yep 3. Adjusting volumes for the median distance uncertainty has raised densities and lowered crossing times by about 3 -- 10 per cent each.

We can also approximate crossing times as how long it takes to cross each half-mass ellipsoid axis at the speed of the associations' 1-dimensional velocity dispersions $\sigma_{v\text{,1D}}$: $t_{\text{cr}}\approx 2r/\sigma_{v\text{,1D}}$ (\citealp{kuhn}; see \S\ref{kinen}), where $r$ is the half-mass radial extent of the ellipsoidal association. Results are 44 -- 100 Myr for UPK 535 and 22 -- 84 Myr for Yep 3, less than or similar to the density-derived crossing times. All these crossing times are reasonable for unbound associations. Furthermore, velocity-dispersion-derived crossing times being roughly equal to association ages imply the associations have expanded since their formation \citep{kuhn}, perhaps partly due to shear forces from the Galactic plane (see Figure \ref{CCmaptoday}). 










\subsection{Colliding Associations} \label{collision}

\begin{figure}
\centering
\includegraphics{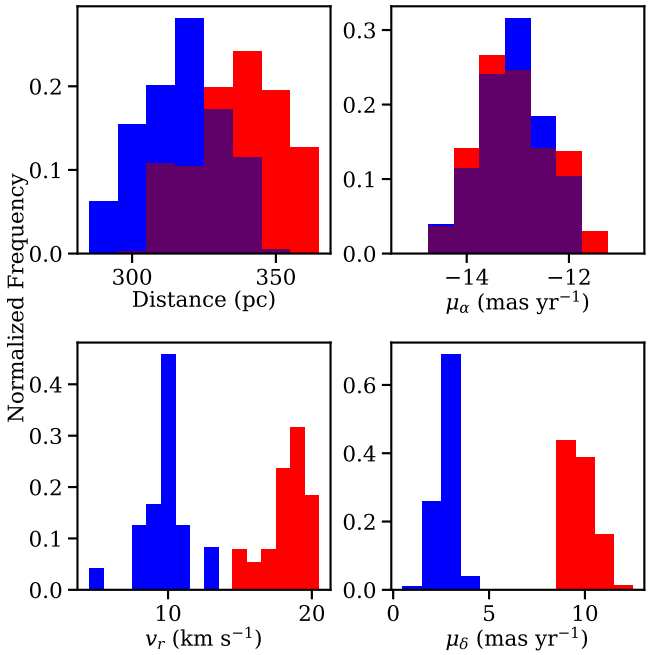}
\caption{Distributions of distances and kinematics for UPK 535 (blue) and Yep 3 (red). The spatially overlapping associations share distances and proper motions in right ascension (purple overlap), but their proper motions in declination and radial velocities diverge, indicating they did not form together. Rather, they recently encountered each other.}
\label{CC_Histograms}
\end{figure}


The observed distributions of distances and kinematics for UPK 535 and Yep 3 are shown in Figure \ref{CC_Histograms}. Their distance distributions overlap, and their proper motions in right ascension $\mu_{\alpha}$ are indistinguishable (see Table \ref{CCpars}). Their motions diverge, however, in proper motion in declination $\mu_{\delta}$ and radial velocity $v_{\text{r}}$. The farther association is moving north and away faster than the nearer association at a relative space velocity of 16.2 $\pm$ 1.4 km s$^{-1}$. Associations UPK 535 and Yep 3 have recently collided.



\setlength\tabcolsep{3pt}
\begin{sidewaystable}[p]
\begin{center}
\vspace{250pt}
\caption{Samples of nominal data for UPK 535 and Yep 3. Star names are available for stars we spectroscopically observed. Full machine-readable tables containing all UPK 535 and Yep 3 stars are available online.}
\label{sample}
\resizebox{\textwidth}{!}{
\begin{tabular}{lcccccccccccccccc}
\hline
 & Star & $\mathit{RA}$ & $\mathit{Dec}$ & Parallax & $\mu_{\alpha}$ & $\mu_{\delta}$ & $v_{\text{r}}$ & G & BP & RP & V & Spectral & Class & Binary & $d$ & Mass  \\
\textit{Gaia} DR2 Source & Name & ($^{\circ}$) & ($^{\circ}$) & (mas) & (mas yr$^{-1}$) & (mas yr$^{-1}$) & (km s$^{-1}$) & (mag) & (mag) & (mag) & (mag) & Type & Uncert.\ & Flag & (pc) & (M$_{\odot}$)  \\
\hline
\multicolumn{17}{c}{UPK 535} \\
\hline
5515311888720004096 & \nodata & 126.3425 & -48.8964 & 3.041 $\pm$ 0.062 & -12.44 $\pm$ 0.10 & 2.77 $\pm$ 0.10 & \nodata  & 16.325 $\pm$ 0.002 & 17.575 $\pm$ 0.013 & 15.183 $\pm$ 0.004 & 17.314 $\pm$ 0.012 & \nodata & \nodata & \nodata & 326.0$^{+6.8}_{-6.5}$ & 0.414 $\pm$ 0.033  \\ 
5323045375612718080 & 2MASS J08280595-4957545 & 127.0248 & -49.9652 & 3.062 $\pm$ 0.016 & -12.913 $\pm$ 0.031 & 2.989 $\pm$ 0.029 & 5.7 $\pm$ 1.7 & 13.334 $\pm$ 0.002 & 13.872 $\pm$ 0.008 & 12.651 $\pm$ 0.006 & 13.639 $\pm$ 0.007 & K3.5V & 1.5 & \nodata & 323.5 $\pm$ 1.7 & 0.798 $\pm$ 0.064  \\ 
5321523750292691968 & TYC 8162-956-1 & 127.5683 & -51.8880 & 3.218 $\pm$ 0.030 & -13.607 $\pm$ 0.054 & 3.268 $\pm$ 0.049 & 10.20 $\pm$ 0.30 & 11.930 $\pm$ 0.002 & 12.262 $\pm$ 0.006 & 11.450 $\pm$ 0.004 & 12.067 $\pm$ 0.020 & F7V & 1.0 & \nodata & 308.0 $\pm$ 2.9 & 1.210 $\pm$ 0.060  \\ 
5322663157872667648 & \nodata & 126.4354 & -50.2127 & 2.994 $\pm$ 0.096 & -12.71 $\pm$ 0.18 & 2.79 $\pm$ 0.19 & \nodata  & 17.428 $\pm$ 0.002 & 18.997 $\pm$ 0.029 & 16.141 $\pm$ 0.004 & 18.724 $\pm$ 0.029 & \nodata & \nodata & \nodata & 331$^{+11}_{-10.}$ & 0.266 $\pm$ 0.021  \\ 
5515843571310261248 & \nodata & 123.8186 & -48.8694 & 3.057 $\pm$ 0.083 & -12.22 $\pm$ 0.15 & 3.87 $\pm$ 0.13 & \nodata  & 16.874 $\pm$ 0.002 & 18.322 $\pm$ 0.020 & 15.653 $\pm$ 0.004 & 18.053 $\pm$ 0.019 & \nodata & \nodata & \nodata & 324.4$^{+9.1}_{-8.6}$ & 0.331 $\pm$ 0.026  \\
\hline
\multicolumn{17}{c}{Yep 3} \\
\hline
5321688200309650432 & TYC 8163-2131-1 & 129.9593 & -51.5401 & 2.876 $\pm$ 0.026 & -12.625 $\pm$ 0.050 & 9.199 $\pm$ 0.053 & -10.15 $\pm$ 0.95 & 11.888 $\pm$ 0.001 & 12.273 $\pm$ 0.004 & 11.346 $\pm$ 0.003 & 12.061 $\pm$ 0.013 & F6V & 3.0 & SB1? & 344.3$^{+3.2}_{-3.1}$ & 1.25 $\pm$ 0.16  \\ 
5321625901794892800 & TYC 8163-1809-1 & 129.9754 & -51.9476 & 2.814 $\pm$ 0.028 & -12.101 $\pm$ 0.050 & 10.556 $\pm$ 0.053 & 12.06 $\pm$ 0.16 & 11.876 $\pm$ 0.000 & 12.163 $\pm$ 0.002 & 11.439 $\pm$ 0.001 & 11.990 $\pm$ 0.005 & F8V & 1.0 & SB1? & 351.8$^{+3.5}_{-3.4}$ & 1.180 $\pm$ 0.059  \\ 
5321634732250617856 & 2MASS J08405718-5145010 & 130.2383 & -51.7503 & 2.813 $\pm$ 0.059 & -12.91 $\pm$ 0.11 & 9.83 $\pm$ 0.13 & \nodata  & 16.323 $\pm$ 0.002 & 17.515 $\pm$ 0.012 & 15.201 $\pm$ 0.004 & 17.255 $\pm$ 0.012 & \nodata & \nodata & \nodata & 352.1$^{+7.5}_{-7.2}$ & 0.428 $\pm$ 0.034  \\ 
5324773773536320512 & CD-50 3593 & 132.3128 & -51.0401 & 2.928 $\pm$ 0.043 & -12.992 $\pm$ 0.077 & 11.482 $\pm$ 0.059 & \nodata  & 9.913 $\pm$ 0.000 & 10.069 $\pm$ 0.001 & 9.678 $\pm$ 0.002 & 9.960 $\pm$ 0.004 & \nodata & \nodata & \nodata & 338.3$^{+5.0}_{-4.9}$ & 1.67 $\pm$ 0.13  \\ 
5317163813045331968 & \nodata & 132.4527 & -54.9563 & 2.789 $\pm$ 0.074 & -12.07 $\pm$ 0.15 & 9.69 $\pm$ 0.14 & \nodata  & 15.803 $\pm$ 0.001 & 16.916 $\pm$ 0.004 & 14.695 $\pm$ 0.002 & 16.659 $\pm$ 0.004 & \nodata & \nodata & \nodata & 355.3$^{+9.6}_{-9.1}$ & 0.454 $\pm$ 0.036  \\
\hline
\end{tabular}}
\end{center}
\end{sidewaystable}
\setlength\tabcolsep{6pt}

\begin{figure}
\centering
\includegraphics{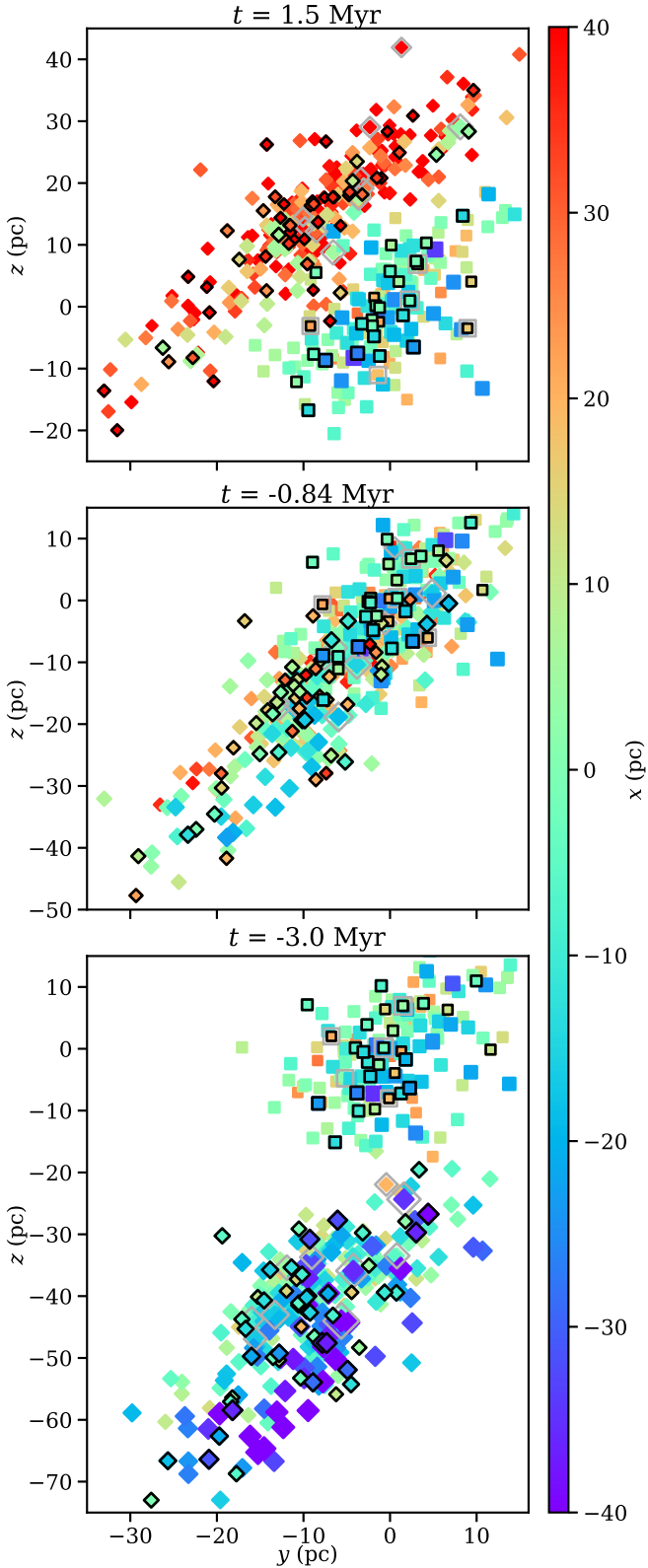}
\caption{Projected positions of UPK 535 (squares) and Yep 3 (diamonds) at three points in time: 3.0 Myr ago (bottom panel), 0.84 Myr ago (middle panel, animated), and 1.5 Myr from now (top panel). $\mathit{RA}$ corresponds roughly with $y$, $\mathit{Dec}$ with $z$, and distance with $x$, all in pc. The coordinates are zeroed on UPK 535. Dark outlines mark stars with spectroscopically measured $v_{\text{r}}$. Outer gray outlines mark potential binaries. Yep 3 moves north and away into UPK 535, reaching closest approach 0.84 $\pm$ 0.03 Myr ago and receding thereafter. This animation and an alternate centred on Yep 3 are available in online resources.}
\label{CollisionPlots}
\end{figure}



\begin{figure}
\centering
\includegraphics{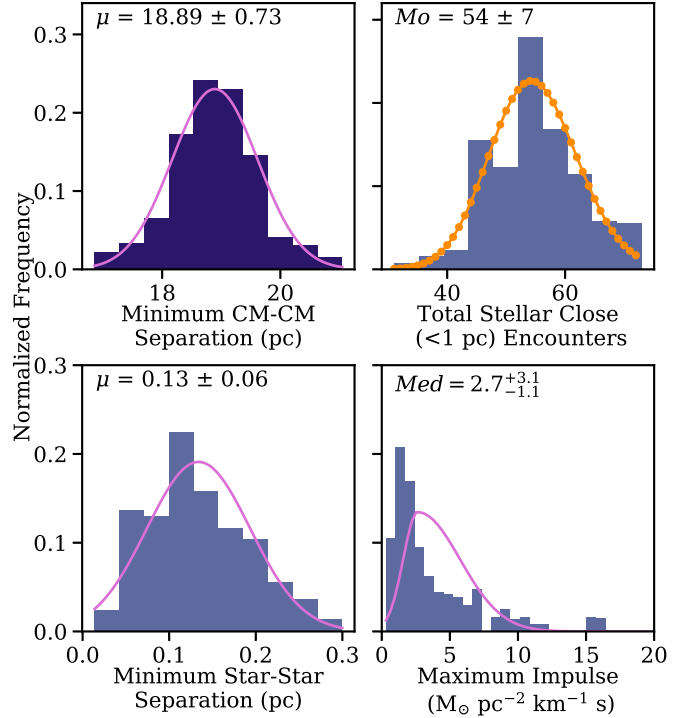}
\caption{Distributions of four parameters across 10,000 trials in our Monte Carlo simulation. Each parameter is fitted with a Gaussian (lavender line) based on mean or median and standard deviation, or a Poisson distribution (orange line) based on mode. The centres of mass (CM) of the two associations (top left) experience an averaged closest approach of 18.89 $\pm$ 0.73 pc about 0.84 $\pm$ 0.03 Myr ago. During the collision, a mode of 54 $\pm$ 7 close ($<$1 pc) stellar encounters occur (top right). The most likely closest approach of any two stars (bottom left) is 0.13 $\pm$ 0.06 pc, or 27,000 $\pm$ 12,000 au, well within the estimated radial size of our Solar Oort cloud. So close an encounter could disrupt debris in a stars' Oort cloud with an impulse-tracing value up to $2.7^{+3.1}_{-1.1}$ M$_{\odot}$ pc$^{-2}$ km$^{-1}$ s (bottom right) and initiate a heavy bombardment of any exoplanets the stars may harbour.}
\label{MonteCarloBest4}
\end{figure}

\begin{figure}
\centering
\includegraphics{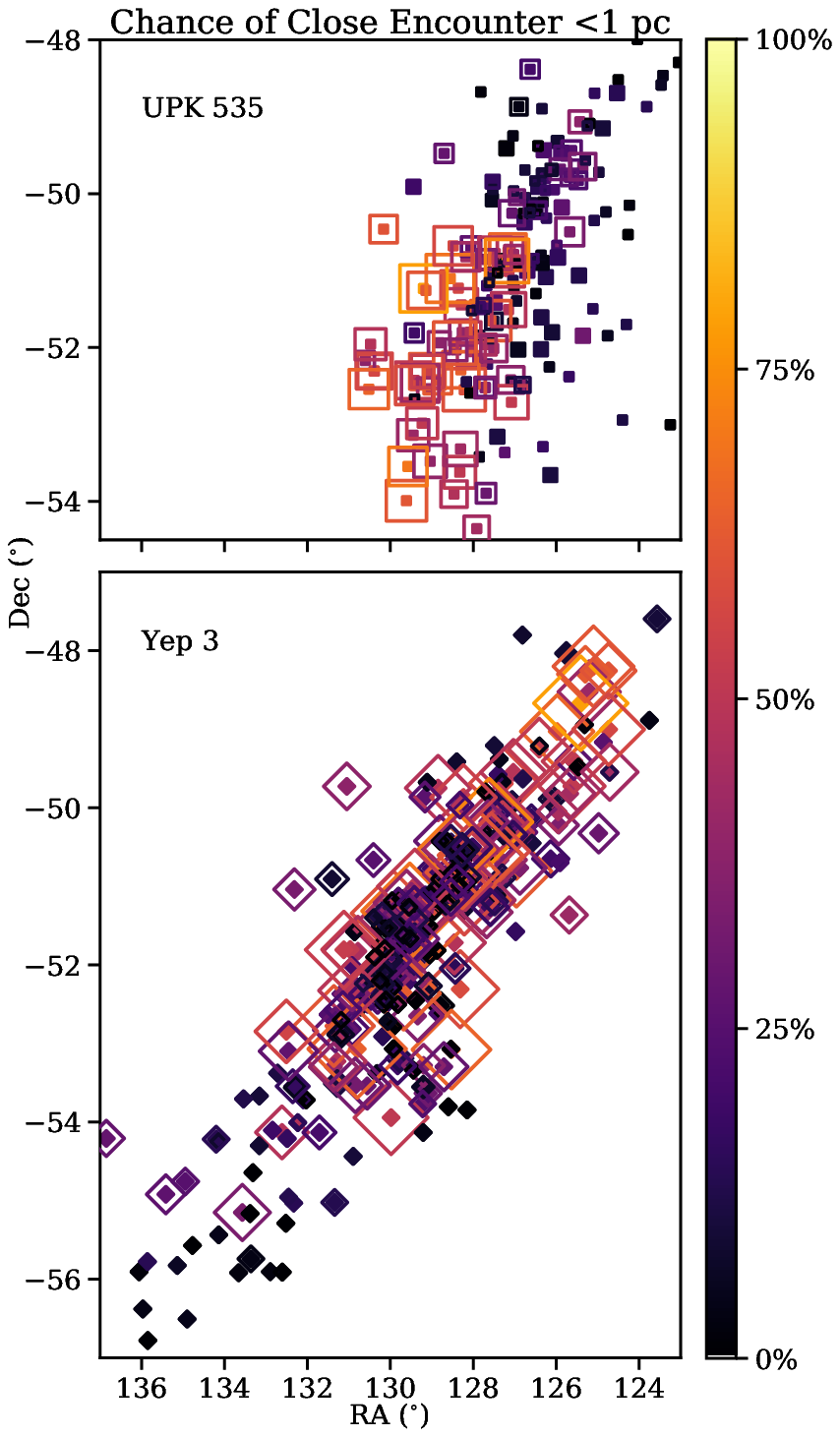}
\caption{Each star's chance of interaction across 10,000 trials. Each symbol's colour represents each star's chance of undergoing a close ($<$1 pc) nonspecific encounter, and each symbol's outline is scaled to represent chance of undergoing more than one close encounter. In UPK 535 (top panel, square symbols), each star's chance of close encounter is strongly dependent on location. Stars in the eastern half of UPK 535 are more likely to encounter Yep 3 stars than are stars in the western half of UPK 535. Chance of close encounter is more spread out in Yep 3 (bottom panel, diamond symbols).}
\label{ChanceofInteraction}
\end{figure}




With median radial extents of 14.6 pc and 16.3 pc, the associations' current centre-of-mass separation of 22.7 $\pm$ 3.8 pc corresponds to 8.2 pc of median radial overlap. To estimate association volume overlap, we scale each association's initial ellipsoid to contain 75 per cent of the association's stars. By a cubic parsec grid calculation, the overlap region includes 26.1 per cent of UPK 535's 75-per-cent-stars ellipsoid volume and 19.3 per cent of Yep 3's. This association overlap is considerable, holding 50 UPK 535 stars and 43 Yep 3 stars (see Figure \ref{anfigrottoday}).




To explore the association collision, we run a 10,000-trial Monte Carlo simulation of the associations' stars' current $xyz$-transformed positions, motions, and uncertainties, linearly extrapolated forward and backward in time (see Figure \ref{CollisionPlots}). Stars without spectroscopically measured $v_{\text{r}}$ and potential binary stars are assigned their association's error-weighted-mean $v_{\text{r}}$ (10.1 km s$^{-1}$ for UPK 535, 19.4 km s$^{-1}$ for Yep 3; see \S3.4 and Table \ref{CG_properties_table.dat}), with uncertainty $\sigma_{v_{\text{r}}}$ set to the association's $v_{\text{r}}$ dispersion (1.6 km s$^{-1}$ for both). We ignore stars' close-encounter path deflections (extremely unlikely; see \citealp{bj2}), accelerations in the gravitational potential of the two associations, and accelerations in the gravitational potential of the Galaxy (see below for effects). Based on the associations' centre-of-mass relative velocity of 16.2 $\pm$ 1.4 km s$^{-1}$ and time-step 0.01 Myr, the collision simulation's spatial resolution is essentially $\sim$0.166 $\pm$ 0.014 pc per time-step. We find that using a finer time-step does not significantly alter our results. Though we do not integrate through gravitational potentials, stellar masses do affect centre-of-mass calculations, so we account for the masses of known potential binaries (see \S\ref{associationprop}) and impose a stellar mass lower limit of 0.08 M$_{\odot}$ on all stars.






In the simulation, the associations' centres of mass experience a mean closest approach of 18.89 $\pm$ 0.73 pc at time 0.84 $\pm$ 0.03 Myr ago (see Figure \ref{MonteCarloBest4} and the middle panel of Figure \ref{CollisionPlots}). Holding the 75 per cent-stars ellipsoids constant and shifting them by their centre-of-mass velocities for 0.84 Myr into the past, we recalculate association volume overlap at time of closest centres-of-mass approach: 23.4 per cent of the volume of UPK 535 and 17.3 per cent of the volume of Yep 3 fall within the overlap region (see 3-D rotating view of UPK 535 and Yep 3 at 0.84 Myr ago, available in online resources). The overlap region at closest centres-of-mass approach encompasses 37 UPK 535 stars and 44 Yep 3 stars.



We track close stellar encounters, defined as closest-approach distances $<$1 pc, within each trial. Measurement uncertainties in distances ($\sim$7.6 pc) and an incomplete sample of radial velocities (21.5 per cent measured) prevent us from predicting specific star-star encounters. However, we can track which stars are most likely to undergo a close encounter with any other-association star. In our Monte Carlo simulation, over the course of 3.15 $\pm$ 0.50 Myr $\ll$ $t_{\text{cr}}$ from the first close stellar encounter to the last, a mode of 54 $\pm$ 7 close stellar encounters occur (see Figure \ref{MonteCarloBest4}). The closest encounter of any two stars is on average 0.13 $\pm$ 0.06 pc $\approx$ 27,000 $\pm$ 12,000 au, a distance well within the estimated extent of the Solar Oort cloud ($\sim$50,000 au).


A star's chance of close encounter is affected by its location within the association (see Figure \ref{ChanceofInteraction}). Stars in UPK 535 have a median 16.9 per cent chance of undergoing a close encounter with a star in Yep 3. The chance of close encounter is stronger for stars in the eastern half of UPK 535 than the western half. Stars in Yep 3 have a median 14.2 per cent chance of close encounter with a star in UPK 535, with chance of close encounter higher in the western two thirds of the association. Two stars in UPK 535 and two in Yep 3 undergo a nonspecific close encounter in $>$70 per cent of trials. These stars undergo multiple close encounters in $\sim$30 per cent of trials. 93.7 per cent of stars in UPK 535 and 84.2 per cent of stars in Yep 3 experience a close encounter in at least one trial. Improved parallaxes with $\sigma_d\sim1$ pc plus additional high-resolution spectroscopic observations with $\sigma_ {v _{\text{r}}}\sim0.1$ km s$^{-1}$ could enable us to predict specific star-star close encounters within 1 -- 2 pc and estimate their effects on the stars' theoretical solar systems.

Ideally we would integrate positions and motions through the gravitational potential of the Galaxy and, if strong enough, the gravitational potentials of the associations themselves. According to \citet{bj2}, who ran both linear and integrated-through-gravitational-potential simulations, our simple linear-motion approach can introduce star-star distance errors $\geq$0.5 pc for up to 17 per cent of stars, with underestimation about twice more frequent than overestimation. This systematic underestimation of star-star distances will systematically inflate the number of close stellar encounters.

\subsection{Impulse} \label{impulse}



It has been theoretically demonstrated that a star passing close to our Sun could dislodge comets from the Oort cloud and send several million comets into the inner Solar System (e.g. \citealp{weiss,yeo,fengbj,bj2}). Long-period comets, because larger and faster-moving than common near-Earth asteroids, may be more likely to inflict extinction-level impacts \citep{weiss,yeo}. Heightened comet fluxes during a close stellar encounter would raise the probability of a catastrophic cometary impact.

The origin of the Sun's Oort Cloud remains mysterious. The Oort Cloud may have evolved over the course of 100 -- 1000 Myr, shaped by planetary migrations, Galactic tidal forces, debris capture, and, in fact, close stellar encounters \citep{hig,zwartO}. The Oort Cloud as we know it could be a unique structure \citep{zwartO}, although the evolution of Oort-like structures around other stars is certainly possible over timescales of 10 -- 200 Myr \citep{zwartO2,zwartO}, and possible microwave evidence of exo-Oort clouds around other stars has been presented \citep{bax,nib}.


If the stars in UPK 535 and Yep 3 do possess $\sim$50,000 au Oort clouds similar to our Sun's, an encounter as close as 27,000 $\pm$ 12,000 au could perturb Oort cloud objects significantly. We estimate stellar influence on cometary motions and induced comet flux through a simple impulse-tracing parameter $Md^{-2}v^{-1}$ \citep{fengbj,bj2}, where $M$ is the mass of the encountering star, $d$ is the star-star separation, and $v$ is the stars' relative space velocity. In the close encounters of our Monte Carlo simulation, where the closest encounter induces the highest impulse just over half of the time, the median maximum impulse parameter any star imparts to another star's Oort cloud comets is $2.7^{+3.1}_{-1.1}$ M$_{\odot}$ pc$^{-2}$ km$^{-1}$ s. Uncertainties are from the first and third quartiles of the asymmetric impulse parameter distribution (see Figure \ref{MonteCarloBest4}). Extrapolating from the Solar-System-based model of \citet{fengbj}, we find the impulse is strong enough to inject a median of 410$^{+560}_{-190}$ of every 1 million Oort Cloud comets into a star's inner solar system. For reference, our own Solar System has an estimated 10$^{11}$ -- 10$^{12}$ Oort Cloud comets, which would entail up to 400 million comets injected \citep{fengbj}. Because travel from the Oort cloud into inner orbits takes time \citep{fengbj}, comet showers may plague the close-encountering stars' exoplanets in a few million years.

A passing star's Oort cloud itself could also sweep through another star's solar system. Additionally, a passing star's tidal tails of asteroids could sweep through another star's solar system millions of years before or after the stars' closest approach \citep{zwartO2}.

If the stars in UPK 535 and Yep 3 do not possess Oort Clouds, it is possible the perturbations from close stellar encounters could spur their creation; a combination of within-cluster stellar interactions and a strong close stellar encounter may have shaped the Sun's Oort Cloud \citep{zwartO}. Perturbation of Kuiper-belt-like structures or asteroids could also occur during close stellar encounters \citep{zwartO2}. Even planets may be slightly shifted: The disturbance of an outer companion can perturb inner planets, perhaps altering orbital distances \citep{wintplanet,longmore} or inclinations \citep{rodet}. Planetary ejections involve stronger, closer encounters within $\sim$1000 au \citep{laugh,vanelt,wright}.


We compare inter-association impulses to intra-association- and field-star-induced impulses. Because the associations are sparse and moving through space together, the separation between stars within each cluster may not change drastically over the course of a few Myr and should hover around $(0.01\text{ per pc}^3)^{-1/3} \approx 5$ pc, which, squared, should dominate the effect of slower velocities and render intra-association impulses overall weaker than the strongest inter-association impulses. However, the field star population (especially in the plane of the Galaxy) is likely denser than each association by at least an order of magnitude. The local field star density near our Sun, for example, is 0.09 stars pc$^{-3}$ \citep{todd}, for typical stellar separation $\sim$2.2 pc. Based on the initial cone search of \textit{Gaia} DR2 stars in the vicinity of UPK 535, the field stars' spread in relative space velocities $\sim$33 km s$^{-1}$ is larger than the relative space velocity 16.2 $\pm$ 1.4 km s$^{-1}$ between UPK 535 and Yep 3, partly offsetting the higher field-star density. For close encounters 1.0 -- 0.1 pc and median stellar mass 0.37 M$_{\odot}$, the field star impulse parameter ranges from 0.01 -- 1.1 M$_{\odot}$ pc$^{-2}$ km$^{-1}$ s, comparable to the inter-association impulse parameter value.




\subsection{Other Colliding Associations}

In the vicinity of UPK 535 and Yep 3 are three other associations, UPK 533, UPK 545, and Pozzo 1 \citep{pozzo,sim,cant}. When we simulate their \textit{Gaia} DR2 positions and motions through time with the same linear approach described above, we find that these associations could also interact with UPK 535, Yep 3, and each other (see Figure \ref{CCmaptoday} and animation in online resources). The centres of mass of UPK 535 and UPK 533 on average reach a closest approach of 23.6 $\pm$ 4.3 pc about 0.94 $\pm$ 0.65 Myr ago, possibly constituting a triple collision for UPK 535 during that time. In the UPK 535-UPK 533 interaction, a mode of 5 $\pm$ 2 close stellar encounters occur. The centres of mass of Yep 3 and UPK 545 on average reach a closest approach of 9.8 $\pm$ 1.7 pc about 0.52 $\pm$ 0.07 Myr from now, during which time a mode of 9 $\pm$ 3 close stellar encounters occur. UPK 533 and Pozzo 1 have an encounter of similar significance to that of UPK 535 and Yep 3. The centres of mass of UPK 533 and Pozzo 1 reach an average closest approach of just 3.4 $\pm$ 2.1 pc about 0.03 $\pm$ 0.14 Myr from now, and a mode of 50 $\pm$ 7 close stellar encounters occur. Minimum stellar separation of any two of their stars is on average 0.13 $\pm$ 0.05 pc $\approx$ 27,000 $\pm$ 10,000 au.

These results for UPK 533, UPK 545, and Pozzo 1 are preliminary, with association radial velocities based on only a handful of available \textit{Gaia} DR2 radial velocities. The stellar membership list of Pozzo 1 is also incomplete. None the less, our tests imply that association interactions are commonplace, at least in the Gum Nebula in the plane of the Galaxy. If so, while star-star interactions may have a limited impact within associations (\citealp{wint}; \S \ref{impulse}), we may have to take increased star-star interactions during association collisions into account when studying how associations, solar system structures, and planets evolve.

\subsection{Kinetic Energy vs.\ Gravitational Potential Energy} \label{kinen}


\begin{figure}
\centering
\includegraphics{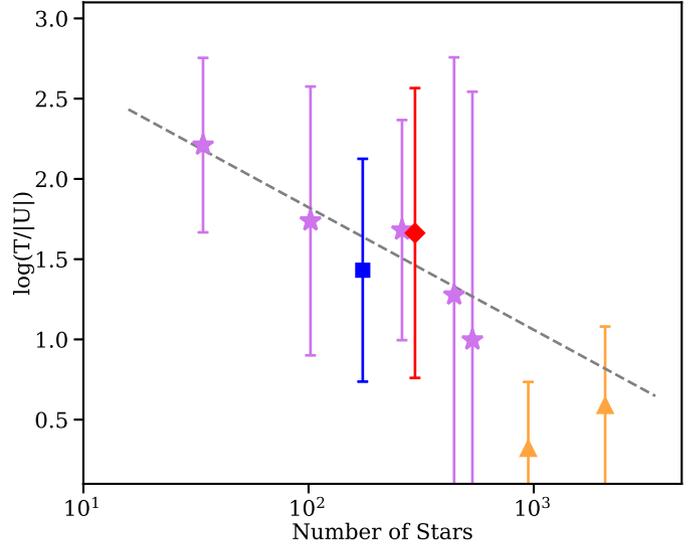}
\caption{Log ratios of associations' kinetic to potential energies are plotted vs.\ the number of stars in each association. We have spectroscopically derived radial velocities for stars in UPK 535 (blue square), Yep 3 (red diamond), and five other associations (lavender stars) throughout the Gum Nebula. For comparison, we include the two subclusters of Cep OB3b from outside the Gum Nebula (orange triangles, \citealp{karnath}). Association energy ratios in the Gum Nebula are consistent with the energy ratios of Cep OB3b. The grey correlation line is an error-weighted fit to the Gum Nebula data.}
\label{EnergiesTU}
\end{figure}

\setlength\tabcolsep{5pt}
\begin{table}
\centering
\caption{Energies and energy log ratios of Gum Nebula associations. Energy log ratio uncertainties for UPK 533, UPK 545, Pozzo 1, and CG 30 exceed 3 dex, so we consider their energies and energy log ratios unmeasured.}
\label{entab}
\begin{tabular}{lcccc}
\hline
Assn. & No. & $T$ & $U$ & $\log(T/|U|)$ \\
Name & Stars & (M$_{\odot}$ km$^2$ s$^{-2}$) & (M$_{\odot}$ km$^2$ s$^{-2}$) & (dex) \\ 
\hline
CG 4 Assn. & 34 & 26 $\pm$ 31 & -0.162 $\pm$ 0.062 & 2.21 $\pm$ 0.54  \\ 
CG 22 Assn. & 102 & 70 $\pm$ 130 & -1.28 $\pm$ 0.20 & 1.74 $\pm$ 0.84  \\ 
Yep 1 & 534 & 300 $\pm$ 1100 & -30.1 $\pm$ 2.3 & 1.0 $\pm$ 1.5  \\ 
Yep 2 & 443 & 260 $\pm$ 890 & -13.8 $\pm$ 1.0 & 1.3 $\pm$ 1.5  \\ 
Yep 3 & 297 & 350 $\pm$ 730 & -7.63 $\pm$ 0.32 & 1.66 $\pm$ 0.90  \\ 
UPK 535 & 174 & 90 $\pm$ 140 & -3.250 $\pm$ 0.099 & 1.43 $\pm$ 0.69  \\ 
Alessi 3 & 260 & 560 $\pm$ 880 & -11.57 $\pm$ 0.27 & 1.68 $\pm$ 0.69  \\
\hline
\end{tabular}
\end{table}
\setlength\tabcolsep{6pt}


An association's ratio of kinetic energy $T$ to absolute value of gravitational energy $|U|$ describes how bound the association is and how quickly it may dissolve \citep{shu,krum}. Systems that are gravitationally bound are expected to obey the virial theorem, $T = -1/2\:U$. A log energy ratio of $\log(T/|U|)$ = -0.3 dex therefore represents a virial, bound, stable star cluster, whereas $\log(T/|U|)>-0.3$ dex is consistent with a supervirial, unbound, expanding stellar association.

To assess the bound or unbound nature of UPK 535 and Yep 3, we compute $\log(T/|U|)$ for them and nine other associations throughout the Gum Nebula, namely the other six associations from our cometary globule search (CG Associations 4, 22, and 30, Yep 1 and 2, and Alessi 3; see \S\ref{ID})  and the three associations UPK 533, UPK 545, and Pozzo 1 in the vicinity of UPK 535 and Yep 3 (see Figure \ref{CCmaptoday}). The kinetic energy $T$ of each association is calculated from stars' velocity dispersion from the association error-weighted-mean velocity:
\begin{equation}
T=\frac{3}{2} M_{\text{tot}} \sigma_{v,\text{1D}}^2.
\end{equation}
\noindent Here, $M_{\text{tot}}$ is the total stellar mass, including masses from binary companions and mass from completing the IMF (see \S\ref{mass}). Because the aberrant radial velocities of single-epoch binary stars can skew kinetic energy towards high values \citep{mathieu,karnath}, the radial velocities of all identified binaries are set to each association's error-weighted mean. With binaries thus controlled, $\sigma_{v\text{,1D}}$ is the 1-dimensional velocity dispersion of the association, derived from the three orthogonal dispersions as $\sigma_{v\text{,1D}}^2 = (\sigma_{v_x}^2+\sigma_{v_y}^2+\sigma_{v_z}^2-\sigma_{v_{\text{r,med}}}^2-\sigma_{v_{\alpha,\text{med}}}^2-\sigma_{v_{\delta,\text{med}}}^2)/3$. Here $\sigma_{v_{\text{r,med}}}$ is the median radial velocity uncertainty of the association or, if the following is smaller, the median radial velocity uncertainty of our whole sample of spectroscopically measured radial velocities in the Gum Nebula, equal to 0.56 km s$^{-1}$. The values $\sigma_{v_{\alpha,\text{med}}}$ and $\sigma_{v_{\delta,\text{med}}}$ are the associations' median proper motion uncertainties in km s$^{-1}$ (see \S\ref{rv}). These three median uncertainties are subtracted in quadrature from the velocity dispersion to avoid artificially inflating $T$ with scatter in motion measurements (see analogous treatment of positional uncertainties in \S\ref{associationprop}).


The gravitational potential energy $U$ is the sum of each star's gravitational potential energy derived from total stellar mass interior to that star:
\begin{equation}
U=-G\frac{ \sum\limits_i \sum\limits_{j, r_j<r_i} m_i m_j}{r_i}.
\end{equation}
\noindent Here, $G$ is the gravitational constant. Quantities $m_i$ and $m_j$ are measured stellar masses, including companion masses of identified potential binaries, times $M_{\text{tot}}/M_{\text{b}}$, where $M_{\text{b}}$ is the sum of all single-star masses and identified binary masses. Multiplying by this factor lets us include masses from unidentified binaries and IMF completion, without altering the identified mass distribution. The value $r$ is a given star's distance from the centre of mass. From this distance, we in quadrature subtract $\sigma_{d_{\text{med}}}$, which is the association's median distance uncertainty or, if the following is smaller, the median distance uncertainty of our whole sample of Gum Nebula stars, equal to 8.5 pc.


The log energy ratio for UPK 535 is 1.43 $\pm$ 0.69 dex, and for Yep 3, 1.66 $\pm$ 0.90 dex (see Table \ref{entab}). Both associations are unbound. Results for five of the nine other associations in the Gum Nebula range from 1.0 to 2.21 dex, also unbound. The remaining four associations (UPK 533, UPK 545, and Pozzo 1, which lack spectroscopic observations, and the sparse CG 30 Association) have large errors in $K$ and $U$ that result in $\log(T/|U|)$ errors $>$3 dex, so we consider their energies and energy log ratios unmeasured and exclude them from Table \ref{entab}. When we calculate $\log(T/|U|)$ in two dimensions instead of three, omitting $x \sim d$, log energy ratios differ by -0.58 -- -0.03 dex from the 3-D values. These are mostly within uncertainties but always lower, suggesting artificial radial elongation and general radial uncertainty systematically elevate our 3-D log energy ratios even after subtracting the median distance uncertainty in quadrature from $r$.



All seven of our measured associations have fewer than 1000 members. The sparser associations tend to have lower $|U|$ than $T$, whereas the more populous associations tend to have more balanced $|U|$ and $T$. An association's $\log(T/|U|)$ appears to be correlated with its star count (see Figure \ref{EnergiesTU}).


We compare these results to the two populous subclusters of Cep OB3b from outside the Gum Nebula, studied by \citet{karnath}. The mean 50-per-cent-binarity $\log(T/|U|)$ for Cep OB3b are 0.58 $\pm$ 0.49 dex for the east subcluster and 0.32 $\pm$ 0.41 dex for the west. Cep OB3b is overall unbound and expanding, but portions of the western subcluster may be bound \citep{karnath}. The Gum Nebula associations' median $\log(T/|U|)$ is 1.66 dex, over 1 dex greater than Cep OB3b's values. However, the Gum Nebula associations' energy vs.\ star count trend remains consistent with Cep OB3b, within 2$\sigma$. Even UPK 535 and Yep 3, despite their recent collision, have energy ratios in line with the other associations and Cep OB3b's. Association collisions may not significantly affect association energies, perhaps due to their brevity or the low densities of the associations in our sample. More comparisons with \textit{Gaia}-observed associations and clusters outside the Gum Nebula are needed to verify this result.










\section{Conclusions} \label{summ}

UPK 535 and Yep 3 in the Gum Nebula are the first observed colliding associations close enough to Earth (318.08 $\pm$ 0.29 pc and 339.54 $\pm$ 0.25 pc, respectively) that their dynamical interaction can be investigated in detail. Our 10,000-trial Monte Carlo simulation reveals the following averaged results:
\begin{itemize}
\item The associations attain a closest centres-of-mass approach of 18.89 $\pm$ 0.73 pc about 0.84 $\pm$ 0.03 Myr ago.
\item 54 $\pm$ 7 close encounters (separation $<$1 pc) occur between stars of UPK 535 and Yep 3.
\item The closest encounter is 0.13 $\pm$ 0.06 pc $\approx$ 27,000 $\pm$ 12,000 au, close enough for stars to sweep through each other's Oort clouds, if the stars possess Oort clouds similar to our Sun's.
\item Two stars in UPK 535 and two in Yep 3 undergo a close encounter in $>$70 per cent of trials, and multiple close encounters in $\sim$30 per cent of trials.
\item The maximum impulse parameter of $2.7^{+3.1}_{-1.1}$ M$_{\odot}$ pc$^{-2}$ km$^{-1}$ s is strong enough to potentially inject 410$^{+560}_{-190}$ of every 1 million Oort Cloud comets into inner orbits and cause heavy bombardment of any exoplanets there, if the stars possess Oort clouds similar to our Sun's.
\item Other Gum Nebula associations (UPK 533, UPK 545, and Pozzo 1) may also be interacting with UPK 535, Yep 3, and each other. Association collisions may be commonplace, at least in the Gum Nebula in the plane of the Galaxy.
\item Gum Nebula association log-ratios of kinetic energy to gravitational potential energy (1.00 -- 2.21 dex) vs.\ star count are consistent with Cep OB3b's, suggesting association collisions may not significantly affect association energies. More robust comparisons with other \textit{Gaia}-observed associations are needed to verify this implication.
\end{itemize}
\noindent The relatively young, nearby associations UPK 535 and Yep 3 provide a case study for association-association interactions and their possible effects on the evolution of solar systems.

\section*{Acknowledgements}

The authors thank Dr. Jao, Dr. Henry, and Dr. Lepine for their early feedback on this discovery. The authors also thank Dr. Karnath, Dr. Matthieu, Dr. Stauffer, Dr. Megeath, and Dr. Mamajek for recommendations on how to analyse association dynamics. This paper was made possible through queued observations on the CHIRON spectrograph on the CTIO/SMARTS telescope on Cerro Tololo, and by the fine people who operate the facility. This research has made use of the VizieR catalogue access tool, CDS, Strasbourg, France (DOI: 10.26093/cds/vizier), with original description published in A\&AS 143, 23. Finally and forever, A. C. Yep thanks her loyal partner R.\ C.\ Marks for his unflagging support.

\section*{Data Availability}



The data underlying this article are available in the article, its online supplementary material, and CHIRON Standards at \url{https://github.com/alexandrayep/CHIRON_Standards}.



\bibliographystyle{mnras}

\bibliography{PaperCC_MNRAS.bib}

\bsp	
\label{lastpage}
\end{document}